\documentclass[epj]{svjour}
\usepackage{graphics}

\usepackage{cite,color}
\usepackage{epsfig}

\newcommand{\nc}{\newcommand}
\nc{\am}[1]{\ensuremath{#1}}

\nc{\ke}{\am{K \to e \nu}}
\nc{\km}{\am{K \to \mu \nu}}
\nc{\kpd}{\am{K_{\pi 2}}}
\nc{\ked}{\am{K_{e2}}}
\nc{\kedg}{\am{K_{e2\gamma}}}
\nc{\ket}{\am{K_{e3}}}
\nc{\kmt}{\am{K_{\mu3}}}

\nc{\kmd}{\am{K_{\mu2}}}
\nc{\erreg}{\am{R_{\gamma}}}
\nc{\kee}{\am{K \to \pi  e \nu}}
\nc{\kmm}{\am{K \to \pi \mu \nu}}
\nc{\nn}{\am{N\!N}}
\nc{\NN}{\am{N\!N}}

\nc{\be}{\begin{equation}}
\nc{\ee}{\end{equation}}
\nc{\ChPT}{$\chi$PT}
\nc{\DAF}{DA\char8NE}
\nc{\GeV}{\mbox{GeV}}
\nc{\ps}{\mbox{ps}}
\nc{\mrad}{\mbox{mrad}}
\nc{\ie}{i.e.}
\nc{\MeV}{\mbox{MeV}}
\nc{\mbo}{\mathversion{bold}}

 \def\f{$\phi$}  \def\ab{$\sim$}  \def\dif{\hbox{d}}
    \def\gam{\gamma}
\def\vp{{\vphantom{$I^{I^i}$}}}

\catcode`@=11 
\newdimen\z@ \z@=0pt
\newskip\z@skip \z@skip=0pt plus0pt minus0pt
\def\m@th{\mathsurround=\z@}
\def\ialign{\everycr{}\tabskip\z@skip\halign} 
\def\eqalign#1{\null\,\vcenter{\openup\jot\m@th
  \ialign{\strut\hfil$\displaystyle{##}$&$\displaystyle{{}##}$\hfil
      \crcr#1\crcr}}\,}
\catcode`@=12 

\renewcommand\Gamma{\char0}
\renewcommand\Delta{\char1}

\begin{document}
\title{\mbo
Precise measurement of $\Gamma(K\to e\nu(\gamma))/\Gamma(K\to \mu\nu(\gamma))$ and
study of $K\to e\nu\gamma$}

\subtitle{The KLOE Collaboration} 

\author{
F.~Ambrosino\inst{3,4} \and
A.~Antonelli\inst{1} \and
M.~Antonelli\inst{1} \and
F.~Archilli\inst{8,9} \and
P.~Beltrame\inst{2} \and
G.~Bencivenni\inst{1} \and
C.~Bini\inst{6,7} \and
C.~Bloise\inst{1} \and
S.~Bocchetta\inst{10,11} \and
F.~Bossi\inst{1} \and
P.~Branchini\inst{11} \and
G.~Capon\inst{1} \and
D.~Capriotti\inst{10}
T.~Capussela\inst{1} \and
F.~Ceradini\inst{10,11} \and
P.~Ciambrone\inst{1} \and
E.~De Lucia\inst{1} \and
A.~De Santis\inst{6,7} \and
P.~De Simone\inst{1} \and
G.~De Zorzi\inst{6,7} \and
A.~Denig\inst{2} \and
A.~Di Domenico\inst{6,7} \and
C.~Di Donato\inst{4} \and
B.~Di Micco\inst{10,11} \and
M.~Dreucci\inst{1} \and
G.~Felici\inst{1} \and
S.~Fiore\inst{6,7} \and
P.~Franzini\inst{6,7} \and
C.~Gatti\inst{1} \and
P.~Gauzzi\inst{6,7} \and
S.~Giovannella\inst{1} \and
E.~Graziani\inst{11} \and
M.~Jacewicz\inst{1}\and
V.~Kulikov\inst{13} \and
G.~Lanfranchi\inst{1} \and
J.~Lee-Franzini\inst{1,12} \and
M.~Martini\inst{1,5} \and
P.~Massarotti\inst{3,4} \and
S.~Meola\inst{3,4} \and
S.~Miscetti\inst{1} \and
M.~Moulson\inst{1} \and
S.~M\"uller\inst{2} \and
F.~Murtas\inst{1} \and
M.~Napolitano\inst{3,4} \and
F.~Nguyen\inst{10,11} \and
M.~Palutan\inst{1} \and
A.~Passeri\inst{11} \and
V.~Patera\inst{1,5} \and
P.~Santangelo\inst{1} \and
B.~Sciascia\inst{1} \and
A.~Sibidanov\inst{1} \and
T.~Spadaro\inst{1} \and
M.~Testa\inst{1} \and
L.~Tortora\inst{11} \and
P.~Valente\inst{7} \and
G.~Venanzoni\inst{1} \and
R.~Versaci\inst{1,5}
}

\institute{
Laboratori Nazionali di Frascati dell'INFN, Frascati, Italy \and %1-LNF
Institut f\"ur Kernphysik, Johannes Gutenberg - Universit\"at Mainz, Germany \and %2-Mainz
Dipartimento di Scienze Fisiche dell'Universit\`a ``Federico II'', Napoli, Italy \and %3-Na
INFN Sezione di Napoli, Napoli, Italy \and %4-NaINFN
Dipartimento di Energetica dell'Universit\`a ``La Sapienza'', Roma, Italy \and %5-Energetica
Dipartimento di Fisica dell'Universit\`a ``La Sapienza'', Roma, Italy \and %6-Rm
INFN Sezione di Roma, Roma, Italy \and %7-RmINFN
Dipartimento di Fisica dell'Universit\`a ``Tor Vergata'', Roma, Italy \and %8-Rm2
INFN Sezione di Roma Tor Vergata, Roma, Italy \and %9-Rm2INFN
Dipartimento di Fisica dell'Universit\`a ``Roma Tre'', Roma, Italy \and %10-Rm3
INFN Sezione di Roma Tre, Roma, Italy \and %11-Rm3INFN
Physics Department, State University of New York, Stony Brook, USA \and %12-NY
Institute for Theoretical and Experimental Physics, Moscow, Russia %13-Moscow
}

\authorrunning{The KLOE Collaboration}

\mail{Mario.Antonelli@lnf.infn.it,\\ Tommaso.Spadaro@lnf.infn.it}

\date{Received: date / Revised version: date}
\abstract{
  We present a precise measurement of the ratio
  $R_K=\Gamma(K\to e\nu(\gamma))/\Gamma(K\to \mu\nu(\gamma))$ and a
  study of the radiative process $K\to e\nu\gamma$, performed with the
  KLOE detector.  The results are based on data
  collected at the Frascati $e^+e^-$ collider \DAF\ for an integrated
  luminosity of 2.2 fb$^{-1}$.
  We find $R_K = (2.493\pm0.025_\mathrm{stat}\pm0.019_\mathrm{syst})\times 10^{-5}$,
  in agreement with the Standard Model expectation. This result is
  used to improve constraints on parameters of the Minimal
  Supersymmetric Standard Model with lepton flavor violation. We also measured
  the differential decay rate $d\Gamma(K\to e\nu\gamma)/dE_\gamma$
  for photon energies $10<E_\gamma<250$ MeV. Results are compared
  with predictions from theory.
} %end of abstract
\titlerunning{Precise measurement of $\Gamma(K\to e\nu(\gamma))/\Gamma(K\to \mu\nu(\gamma))$ and study of $K\to e\nu\gamma$}
\maketitle
\section{Introduction}
\label{intro}
The decay $K^\pm\!\to e^\pm\nu$ is strongly suppressed,
$\sim$few$\times$10$^{-5}$, because of conservation of angular
momentum and the vector structure of the charged weak current. It
therefore offers the possibility of detecting minute contributions from
physics beyond the Standard Model (SM). This is particularly true of the
ratio $R_K=\Gamma(\ke)/\Gamma(\km)$ which, in the SM, is
calculable without hadronic uncertainties \cite{marci,Cirigliano:2007xi}.
Physics beyond the SM, for example mul\-ti\--Higgs effects inducing an effective
pseudo-scalar interaction, can change the value of $R_K$.
It has been shown in Ref. \citen{masiero} that deviations of $R_K$ of
up to {\em a few percent} are possible in minimal
supersymmetric extensions of the SM (MSSM) with non vanishing $e$-$\tau$
scalar lepton mixing. To obtain
accurate predictions, the radiative process $K\to e \nu \gamma$ (\kedg)
must be included. In \kedg, photons can be produced via
internal-bremsstrahlung (IB) or  direct-emission (DE), the latter being
dependent on the hadronic structure.  Interference among the two
processes  is negligible \cite{bijnens}.  The DE contribution to the
total width is approximately equal to that of IB \cite{bijnens} but is
presently known with a 15\% fractional accuracy \cite{Heintze:1977kk}.

$R_K$ is {\em defined} to be inclusive of IB, ignoring however
DE contributions.  A recent calculation \cite{Cirigliano:2007xi},
which includes order $e^2p^4$ corrections in chiral perturbation
theory (\ChPT), gives:
\begin{equation}
\label{eq:rksm}
 R_K = (2.477\pm0.001)\times 10^{-5}.
\end{equation}
$R_K$ is not directly measurable, since
IB cannot be distinguished from DE on an event-by-event basis.
Therefore, in order to compare data with the SM prediction at the
percent level or better, one has to be careful with the DE part.\footnote{
 The same arguments
 apply in principle to $\Gamma$(\km). However, there is no helicity
 suppression in this case. IB must be included and DE can be safely
 neglected.}

\begin{figure}[ht!]\centering
  \includegraphics[width=0.7\linewidth]{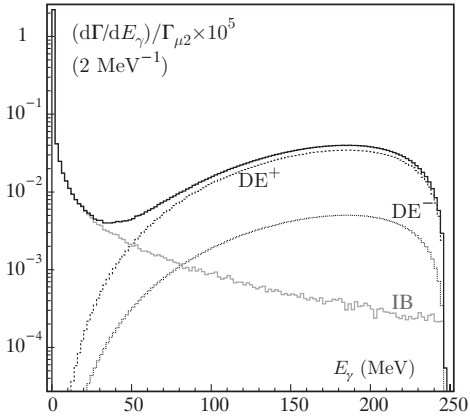}
  \caption{CM photon spectrum for \kedg\ decay. Inner brem\-sstrahlung (IB) and positive and
    negative helicity direct emission (DE$^+$ and DE$^-$) contributions are also shown.}
  \label{fig:egamspectra}
\end{figure}

DE can proceed through vector and axial-vector transitions, with
effective coupling $V$ and $A$, respectively:
\begin{equation}
\eqalign{
{{\rm d}^2 \Gamma(\kedg,{\rm DE})\over{\rm d}x\,{ d}y}&=
\frac{G_F^2\left|\sin\theta_{\mathrm C}\right|^2\alpha_{\rm em} M_K^5}{64\pi^2}\times\cr &\kern-2cm\left[(V+A)^2f_{\mathrm{DE}^+}(x,y)+(V-A)^2f_{\mathrm{DE}^-}(x,y)\right],\cr}
\label{eq:dGdeDE}
\end{equation}
where $G_F$ is the Fermi coupling, $\theta_{\rm C}$ is the Cabibbo
angle \cite{Cabibbo:1963yz}, $x=2E_\gamma/M_K$, $y=2E_e/M_K$ are the
dimensionless photon and electron energies in the kaon rest frame
(both lying between 0 and 1), and 
\begin{equation}
\eqalign{
f_{\mathrm{DE}^+}(x,y)&=(x+y-1)^2(1-x),\cr
f_{\mathrm{DE}^-}(x,y)&=(1-y)^2(1-x).\cr}
\end{equation}
Terms proportional to $(m_e/M_K)^2$ are neglected.
The photon energy spectrum in the CM is shown in Fig. \ref{fig:egamspectra} with its
IB, DE$^+$, and DE$^-$ contributions.\footnote{``+'' and ``$-$'' refer to the photon helicity.}
The DE terms are evaluated with constant $V,\ A$ coupling and calculated in \ChPT\ at
${\mathcal O}(p^4)$ \cite{bijnens}.
%The IB and DE spectra are quite different. 
We define the rate $R_{10}$ as:
\be
R_{10}=\Gamma(\ke(\gamma),\ E_\gamma<10\mathrm{\ MeV})/\Gamma(\km).
\label{eq:R10}
\ee
Evaluating the IB spectrum to ${\mathcal O}(\alpha_{\rm em})$
with resummation of leading logarithms, $R_{10}$ includes
$93.57\pm0.07\%$ of the IB,
\be
R_{10}=R_K\times(0.9357\pm0.0007).
\label{eq:R10new}
\ee
The DE contribution in this range is expected to be negligible.
$R_{10}$ is measured without photon detection. 
Some small contribution of DE is present in the selected sample.
In particular, DE decays have some overlap
with the IB emission at high $p_e$.
We have also measured the differential width
\be
\frac{\mathrm{d}R_\gamma}{\mathrm{d}E_\gamma}=
\frac{1}{\Gamma(\km)}\frac{\mathrm{d}\Gamma(\ke\gamma)}{\mathrm{d}E_\gamma},
\label{eq:rgamma}
\ee
 for $E_\gamma\!>10$ MeV and $p_e\!>200$ MeV
requiring photon detection, both to test \ChPT\ predictions for the
DE terms and to reduce possible systematic uncertainties on the
$R_{10}$ measurement.

\section{DA\char8NE and KLOE}
\label{sec:detector}
\DAF, the Frascati $\phi$ factory, is an $e^{+}e^{-}$ collider
operated at a total energy $\sqrt{s}=m_\phi$\ab1.02 GeV. $\phi$ mesons are
 produced,
essentially at rest, with a visible cross section of $\sim$ 3.1 $\mu$b
and decay into $K^+K^-$ pairs with a BR of $\sim 49$\%.
 During 2001-2005 KLOE collected an integrated luminosity of about
2.2 fb$^{-1}$,  corresponding to $\sim$3.3 billion of $K^+K^-$ pairs.
Kaons have a momentum of \ab100 MeV corresponding to a velocity $\beta_K$\ab0.2.
The mean kaon path is $\lambda_K$\ab90 cm. Observation of a $K^\pm$ meson signals, 
or tags, the presence of a $K^\mp$ meson.
Kaon production and decay are studied with the KLOE detector, consisting essentially of a drift
chamber, DC, surrounded by an
electromagnetic calorimeter, EMC. A superconducting coil provides a 0.52 T magnetic field.

The DC, see Ref. \citen{KLOE:DC}, is a cylinder of 4 m in diameter
and 3.3 m in length. It contains 12,582 drift cells arranged in 58 stereo
layers uniformly filling the sensitive volume.
 The momentum resolution for tracks at large polar
 angle is $\sigma(p_\bot/p_\bot)\!\leq\!0.4$\%.

\begin{figure}[ht]\centering
  \includegraphics[width=0.6\linewidth]{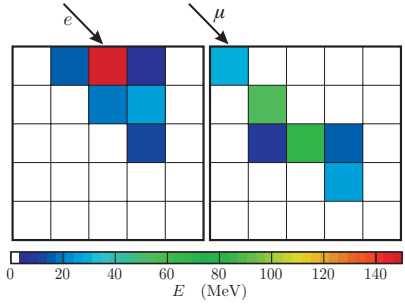}
  \caption{Energy deposit pattern in the EMC cells for a 200 MeV electron (left) and a muon
    (right) from two $K_L\to \pi \ell \nu$ events.}
  \label{ke2:display}
\end{figure}

The EMC is a lead/scintillating-fiber sampling calori\-meter \cite{KLOE:EmC}
consisting of a barrel and two endcaps, with good
energy resolution, $\sigma_{E}/E \sim 5.7\%/\sqrt{\rm{E(GeV)}}$, and excellent
time resolution, $\sigma_{T} =$ 54 ps$/\sqrt{\rm{E(GeV)}} \oplus 140$ ps.
The EMC provides also particle identification, based on the pattern of
energy deposits in the EMC cells.  An example of the difference between
electron and muon patterns is shown in Fig. \ref{ke2:display}.

The trigger \cite{KLOE:trig} uses both EMC and DC information.
Two energy deposits above threshold ($E>50$ for barrel and
$>150$ MeV for endcaps) are required for the EMC trigger.
The DC trigger is based on wire hit multiplicity.
The logical OR of EMC and DC triggers is used for the measurement
presented. The trigger efficiency is evaluated from data.

Cosmic-ray rejection is performed by the trigger hardware. Residual cosmic ray
and machine background events are removed by an offline 
software filter using calorimeter information before track reconstruction.

The detector response is obtained by means of the KLOE Monte
Carlo (MC) simulation program Geanfi, Ref. \citen{nimoffline}.  Changes in machine
parameters and background conditions are simulated on a run-by-run
basis in order to properly take into account the induced effects.% of changes in machine operation.

The MC samples used for this analysis correspond to integrated
luminosities of 4.4 fb$^{-1}$ for the main $K^\pm$ decay modes and of
500 fb$^{-1}$ for decays with BR's less than 10$^{-4}$.  The
effects of initial- and final-state radiation are included in the
simulation at the event generator level \cite{nimoffline,teorico}.
For \kedg\ events, the IB component is described at
${\mathcal O}(e^2)$ in\-clu\-ding resummation of leading
logarithms \cite{teorico}, while the DE component is described with
\ChPT\ at ${\mathcal O}(e^2p^4)$ \cite{bijnens}.
Unless otherwise specified, when comparing data with simulation
we rescale MC samples to an integrated luminosity of 2.2 fb$^{-1}$,
assume the SM value for $R_K$, and use the theoretical prediction for the
DE/IB fraction.

\section{Selection of leptonic kaon decays}
\label{sec:method}
\begin{figure}[ht]\centering
  \includegraphics[width=0.9\linewidth]{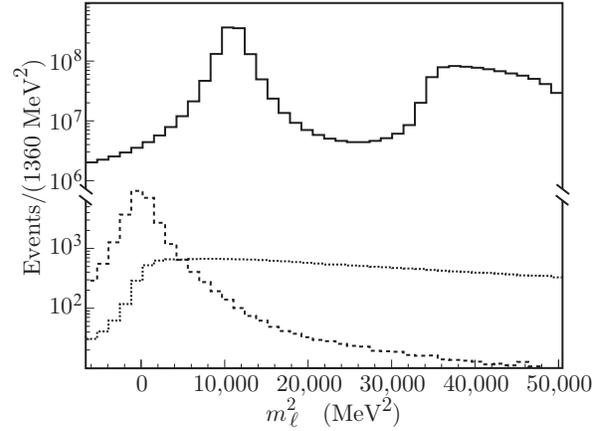}
  \caption{MC distribution of $m_\ell^2$, solid
    line. The contribution of \kedg\ with $E_\gamma< 10$ MeV
    ($>10$ MeV) is shown by the dashed (dotted) lines.}
  \label{fig:mlep}
\end{figure}

$K^\pm$ decays are signaled by the observation of two tracks
with the following conditions. One track must originate at the interaction point (IP) and have
 momentum in the interval $\{70,\:130\}$ MeV, consistent with being a kaon from \f-decay. The
second track must originate at the end of the previous track and have momentum larger than that
of the kaon, with the same charge. The second track is taken as a decay product of the kaon. The
point of closest approach of the two tracks is taken as the kaon decay point D and must satisfy
40$<r_{\rm D}<$150 cm, $|z_{\rm D}|<$80 cm. The geometrical acceptance with these conditions
is \ab56\%, while the decay point reconstruction efficiency is \ab51\%. From the measured kaon
and decay particle momenta, ${\bf p}_K$ and ${\bf p}_{\rm d}$, we compute the squared mass
$m_\ell^2$ of the lepton for the decay $K\to\ell\nu$ assuming zero missing mass:
\be
m_\ell^2=\left(E_K-\left|{\bf p}_K-{\bf p}_{\rm d}\right|\right)^2-{\bf p}_{\rm d}^2{\mbox .}
\label{eq5}
\ee
The distribution of $m_\ell^2$ is shown in Fig. \ref{fig:mlep}, upper curve, from MC simulation.
The muon peak is quite evident, higher masses corresponding to non leptonic and semileptonic
decays. No signal of the $K\to e\nu$ (\ked) decay is vi\-si\-ble.
The very large background around zero mass
is the tail of the \km\ (\kmd) peak, due to poor measurements of $p_K$,
$p_{\rm d}$ or the decay angle, $\alpha_{K\rm d}$.
The expected signal from \kedg\ is also shown in Fig. \ref{fig:mlep}, lower curves, separately
for $E_\gamma>$10 and $<$10 MeV. The expected number of \ked\
decays in the sample is \ab30,000. A background rejection of at least 1000 is necessary, to
obtain a 1\% precision measurement of $\Gamma(\ked)$, with an efficiency of \ab30\%.

The kinematics of the two-body decay $\phi\to K^+K^-$ provides an
additional measurement of $p_K$.
The kaon momentum at the IP is computed from its direction at the IP and the known value of the
$\phi$ 4-mo\-men\-tum.\footnote{The average value of $\phi$ 4-mo\-men\-tum is determined on a
run-by-run basis from Bhabha events, while event-by-event fluctuations
are dominated by the beam energy spread.} The computed value is extrapolated to the decay
point D, accounting for $K$ energy losses in the material traversed. These
are relevant, since the kaon velocity is $\sim0.2$. The material amount traversed
has been determined to within 1\%, thus reducing its contribution
to the momentum resolution to below 0.5 \MeV. The total resolution of
the measurement is $\sim1$ \MeV, comparable with that from track
reconstruction.
We require the two $p_K$ determinations to agree within 5 MeV.

Further cuts are applied to the daughter track. Resolution of track
parameters is improved by rejecting badly reconstructed tracks, i.e.,
with $\chi^2({\rm track\ fit)}/\mathrm{ndf}\!>\!7.5$. 
Events with poorly determined decay angles are mostly due to tracks with improper left-right 
assignment in the reconstruction of the DC hits. 
This happens often when a large majority of the hits associated to the daughter track are on
a single stereo view.
These events are removed by a cut on the
the fractional difference of the number of hits on each stereo view.

Finally, using the expected errors on $p_K$ and
$p_{\rm d}$ from tracking, we compute event by event the error on $m_\ell^2$,
$\delta m_\ell^2$. 
The distribution of $\delta m_\ell^2$ depends slightly on the opening angle $\alpha_{Kd}$,
which in turn has different distribution for \ked\ and \kmd.
Events with large value of $\delta m_\ell^2$ are rejected: $\delta m_\ell^2<\delta_\mathrm{max}$, with
$\delta_\mathrm{max}$ defined as a function of $\alpha_{Kd}$,
to equalize the losses due to this cut for \ked\ and \kmd.

\begin{figure}[ht]\centering
  \includegraphics[width=0.8\linewidth]{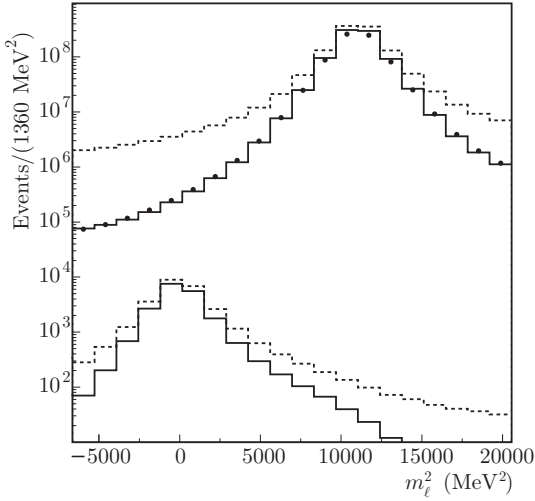}
  \caption{$m_\ell^2$ spectrum before (dashed) and
    after (solid) quality cuts for MC \kmd\ (upper plots)
and \ked\ with $E_\gamma<10$ MeV (lower plots). Black dots represent data after quality
    cuts.}
  \label{fig:mlep_2}
\end{figure}

The effect of quality cuts on $m_\ell^2$ resolution is shown
in Fig. \ref{fig:mlep_2}.  The background in the \ked\ signal region is
effectively reduced by more than one order of magnitude with an
efficiency of $\sim$70\% for both \ked\ and \kmd.

Information from the EMC is also used to improve background rejection. To
this purpose, we extrapolate the secondary track to the EMC surface
and associate it to a nearby EMC cluster.  This requirement produces a
signal loss of about 8\%.

Energy distribution and position along the shower axis of all cells
associated to the cluster allow for $e/\mu$ particle identification.
For electrons, the cluster energy $E_\mathrm{cl}$ is a
measurement of the particle momentum $p_d$, so that
$E_\mathrm{cl}/p_d$ peaks around 1, while for muons
$E_\mathrm{cl}/p_d$ is on average smaller than 1.
Moreover, electron clusters can also be distinguished from $\mu$
(or $\pi$) clusters by exploiting the granularity of the EMC:
electrons shower and deposit their energy mainly in the first plane of
EMC, while muons behave like minimum ionizing particles in the first
plane and deposit a sizeable fraction of their kinetic energy from the
third plane onward, when they are slowed down to rest (Bragg's peak),
see Fig. \ref{ke2:display}.

\begin{figure}[ht]\centering
  \includegraphics[width=0.7\linewidth]{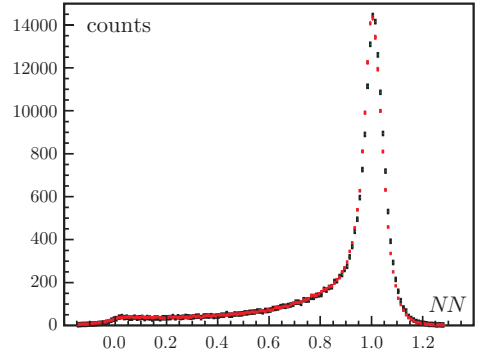}
  \caption{Neural-network output, \nn, for electrons of a $K_L\to \pi e \nu$
   sample from data (black) and MC (red).}
  \label{ke2:pidNN}
\end{figure}

\begin{figure}[ht]\centering
  \includegraphics[width=0.8\linewidth]{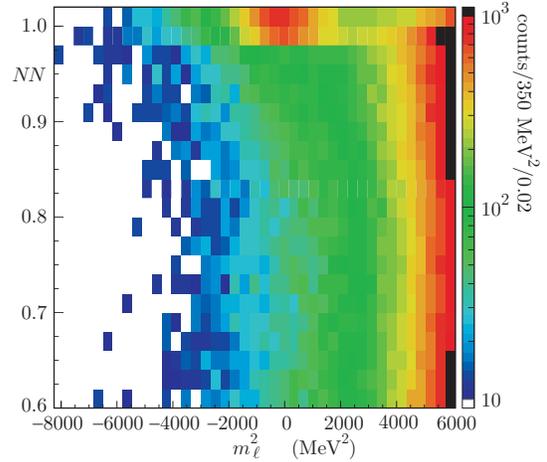}
  \caption{Data density in the \nn, $m_\ell^2$ plane.}
  \label{ke2:pidNNmlep2}
\end{figure}

All useful information about shower profile and total energy deposition are
combined with a 12-25-20-1 structure neural network trained on
$K_L\to \pi \ell \nu$ and \kmd\ data, taking into
account variations of the EMC response with momentum and impact angle
on the calorimeter. The distribution of the neural network output, \nn, for a
sample of  $K_L\to \pi e \nu$ events is shown in Fig. \ref{ke2:pidNN},
for data and MC. Additional separation has been obtained using time
of flight information.
The data distribution of \nn\ as function of $m_\ell^2$ is
shown in Fig. \ref{ke2:pidNNmlep2}. A clear \ke\ signal can be seen
at $m_\ell^2\sim0$ and $\nn\sim1$.

Some $32\%$ of the events with a $K$ decay in the fiducial volume,
have a reconstructed kink matching the required quality criteria {\em and} an EMC cluster
associated to the lepton track; this holds for both \ked\ and
\kmd. In the selected sample, the contamination from $K$ decays other than $K_{\ell2}$ is negligible, as
evaluated from MC.  
$R_{10}$, Eq. \ref{eq:R10}, is obtained without requiring the presence of the radiated photon. 
The number of $\ke(\gamma)$, is determined with a binned
likelihood fit to the two-dimensional \nn vs $m_\ell^2$ distribution.
Distribution shapes for signal and \kmd\ background are taken from MC; the normalization
factors for the two components are the only fit parameters. The fit has been performed
in the region $-3700<m_\ell^2<6100$ MeV$^2$ and $\nn>0.86$.
The fit region accepts $\sim90\%$ of $\ke(\gamma)$ events with $E_\gamma<10$ MeV, as evaluated from MC.
A small fraction of fitted $\ke(\gamma)$ events have $E_\gamma>10$ MeV: the value of this
``contamination'', $f_{\rm DE}$, is fixed in the fit to the expectation from simulation,
$f_\mathrm{DE} = 10.2\%$. A systematic error related to this assumption is discussed in Sect. \ref{sec:syst}.

\begin{figure*}[ht]\centering
    \includegraphics[width=0.4\textwidth]{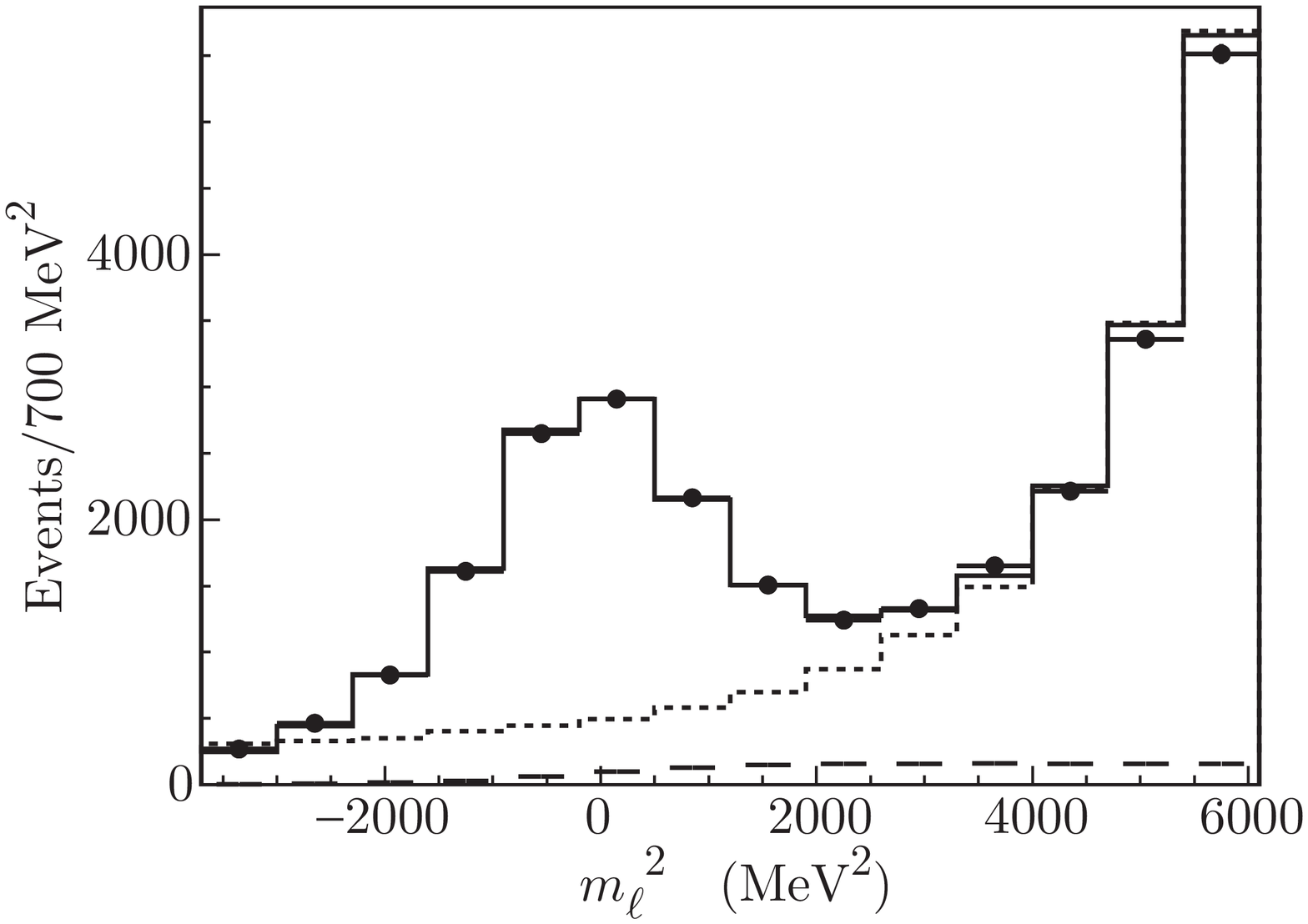}\kern1cm
    \includegraphics[width=0.4\textwidth]{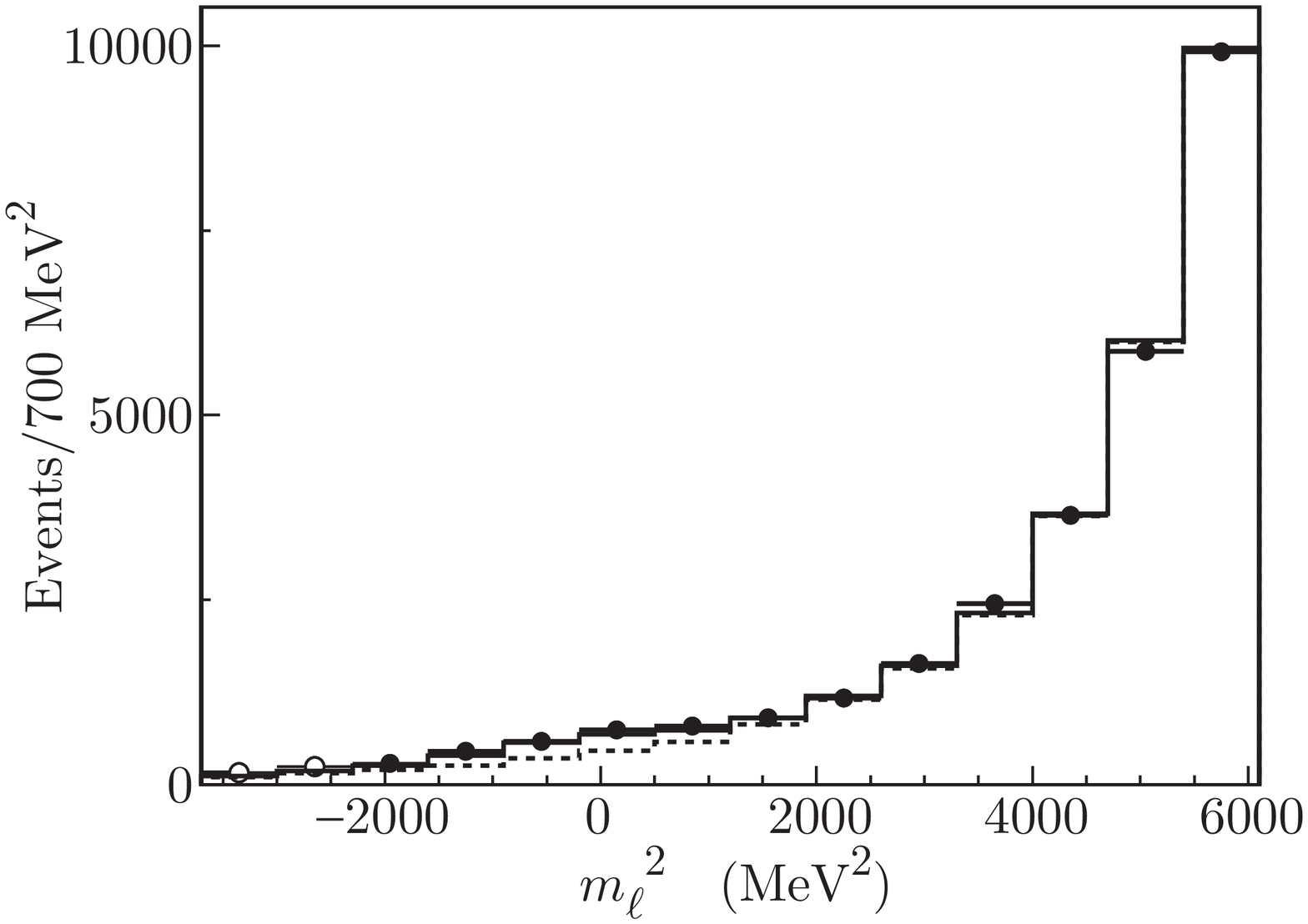}
  \caption{Fit projections onto the $m_\ell^2$ axis for $\nn>0.98$ (left) and $\nn<0.98$ (right), for data (black dots), MC fit (solid line), and  \kmd\ background (dotted line). The contribution from \ked\ events
with $E_\gamma>10$ MeV is visible in the left panel (dashed line).}
  \label{fig:fitke2}
\end{figure*}

We count 7064$\pm$102 $K^+\to e^+\nu(\gamma)$ events and 6750$\pm$ 101 $K^-\to e^-\bar{\nu}(\gamma)$,
89.8\% of which have  $E_\gamma<10$ MeV.
The signal-to-background correlation is $\sim 20\%$ and the
$\chi^2/\mathrm{ndf}$ is 113/112 for $K^+$ and 140/112 for $K^-$.\footnote{The
$\chi^2/\mathrm{ndf}$ of the $K^-$ fit improves to 114/98 for a fit range $\nn>0.88$,
with negligible difference in the measured value for $R_{10}$.}
Fig. \ref{fig:fitke2} shows the sum of fit results for $K^+$ and
$K^-$ projected onto the $m_\ell^2$ axis in a signal
($\nn>0.98$) and a background ($\nn<0.98$) region.  The
residual contribution of \kedg\ events with $E_\gamma>10$ MeV is also shown.

The number of \kmd\ events is obtained from a fit to the $m_\ell^2$ distribution.  The fraction of
background events under the muon peak is estimated from MC to be less
than one per mil. We count $2.878\times10^8$ ($2.742\times10^8$)
$K^+\to \mu^+\nu(\gamma)$ ($K^-\to \mu^-\bar{\nu}(\gamma)$) events.  The difference between $K^+$ and
$K^-$ counts is due to $K^-$ nuclear interactions in the material traversed.

\subsection{\mbo \kedg\ event counting}
\label{sec:kedg}
\begin{figure}[ht]\centering
  \includegraphics[width=0.7\linewidth]{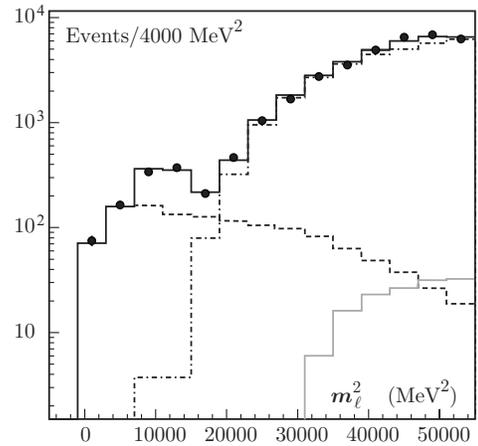}
  \caption{$m_\ell^2$ distribution for data (black dots) and MC (solid line) for events with a detected 
    photon. MC \kedg\ events with $p_e\!<\!$ 200 MeV (gray), $p_e\!>\!$ 200 MeV (dashed) and \ket\ events
    (dot-dashed) are shown separately.}
  \label{fig:mlep_g}
\end{figure}

In order to study \kedg\ decays, we apply the same selection criteria as for \ked, but a tighter PID cut,
$\nn>0.98$. We also require one and only one photon in time with the $K$ decay.
Photons are identified by selecting a cluster with energy greater than
20 MeV. This requirement reduces machine background
and suppresses most of the IB events, leaving a sample dominated by direct emission process (DE).
Moreover, the difference between the photon and the electron measured time of flight has to lie within two 
standard deviations from its expected value.
The fraction of signal events satisfying all of these additional requests is $\sim$ 25\%.
The $m_\ell^2$ distribution for the selected events is shown in Fig. \ref{fig:mlep_g} for data and MC.
\kedg\ decays with $p_e>200$ MeV and $p_e<200$ MeV are shown separately. 
The high-momentum component is dominated by DE$^+$ process,
DE$^-$ accounting for 2\% only (Eq. \ref{eq:dGdeDE}), and is the only relevant for the systematic 
related to the $R_{10}$ measurement: 
high $p_e$ values correspond to low values of $m_\ell^2$ where the \ked\ signal lies.
The low-momentum component, with contributions from both DE$^+$ and DE$^-$ processes,
is completely overwhelmed by \ket\ events with one undetected photon from $\pi^0$ decay.

Further rejection of \ket\ events is provided by kinematics. 
The photon energy in the laboratoty frame, $E_\gamma(\mathrm{lab})$, can be calculated for \kedg\ decays 
from the measured photon direction, the kaon momentum ${\rm p}_K$ and the electron momentum ${\rm p}_e$, 
with a resolution of $\sim12$ MeV. 
The resolution on $\Delta E=E_\gamma(\mathrm{lab})-E_{\gamma,\:{\rm EMC}}$ is that of the calorimeter,
$\sigma\sim30$ MeV for $E_\gamma(\mathrm{lab}) = 200$ MeV.
The number of \kedg\ events is found from a binned likelihood fit in the 
$\Delta E/\sigma$-$m_\ell^2$ plane. This provides a better signal to noise figure, 
compared to using cuts on $\Delta E$ and $m_\ell^2$.
Distribution shapes for signal and \kmd\ and \ket\ backgrounds are taken from MC. 
The amounts of the three components are the fit parameters.

\begin{figure}[ht]\centering
   \includegraphics[width=0.35\textwidth]{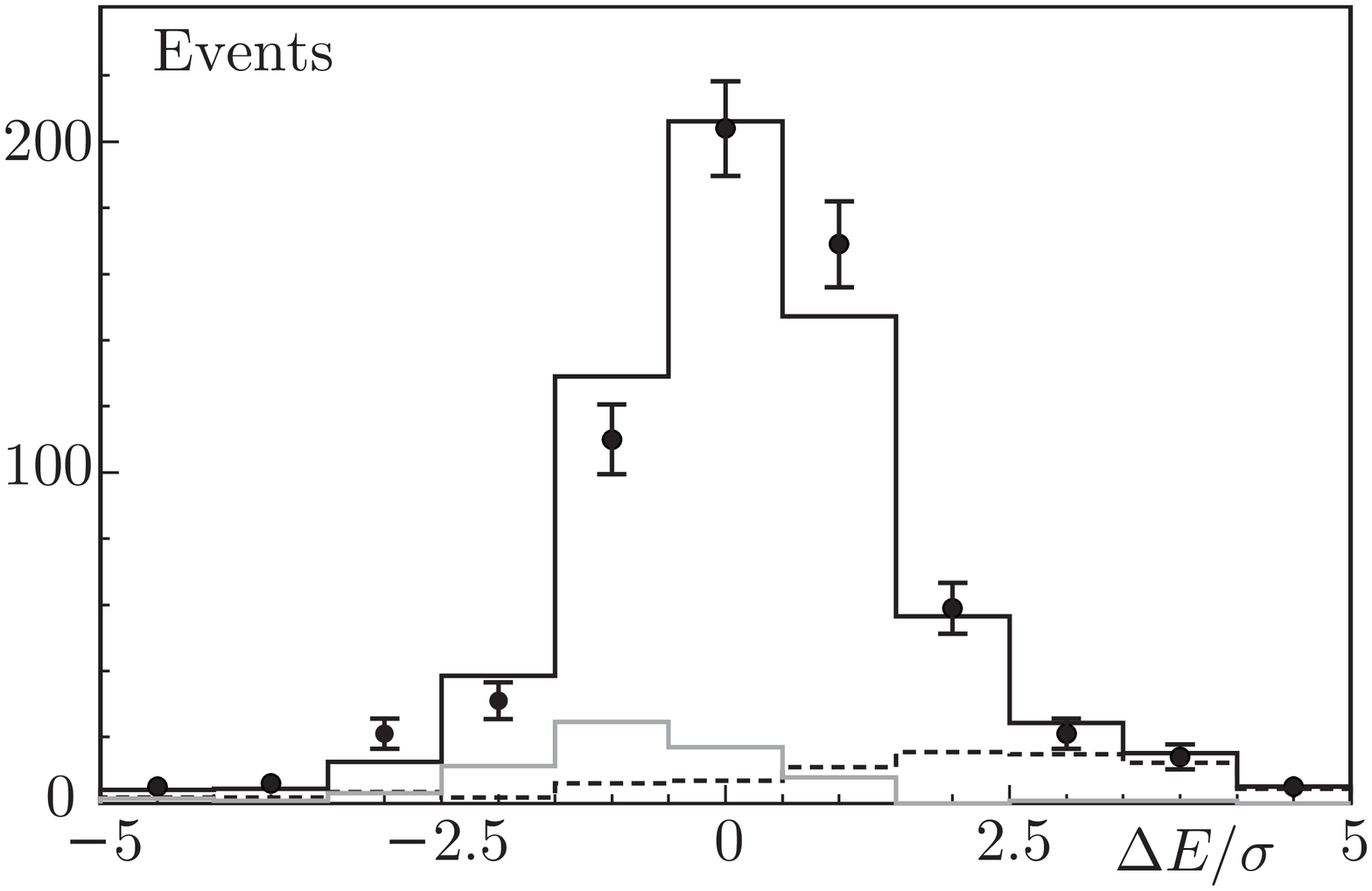}\\ \includegraphics[width=0.35\textwidth]{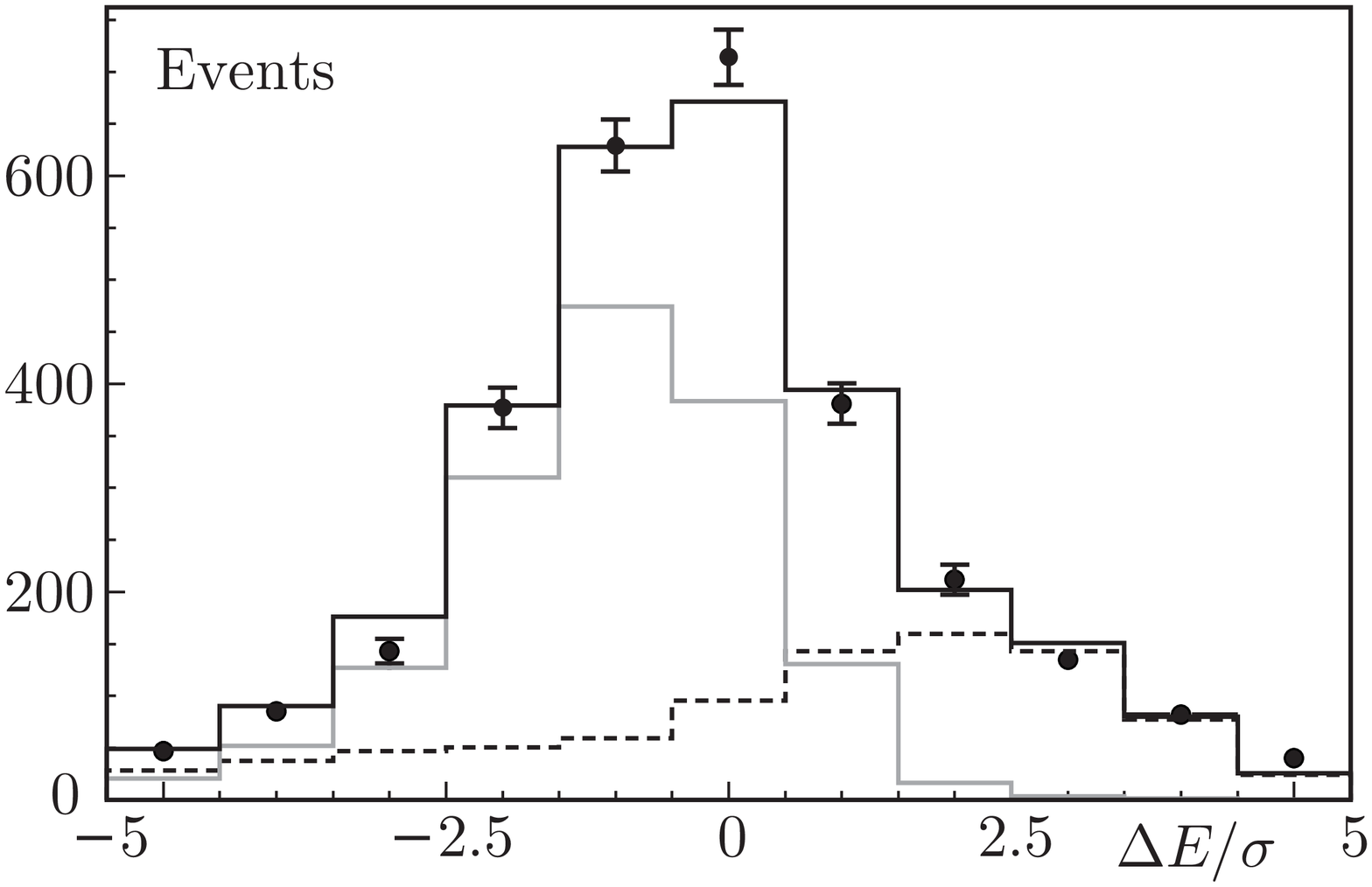}
  \caption{\kedg\ fit projections onto the $\Delta E/\sigma(\Delta E)$ axis for the signal region 
  (as defined in the text, top) and background region (bottom) for data (black dots), MC fit (solid line),
  \kmd\ (dashed line) and \ket\ (gray line). All $E_\gamma$ bins are added.}
  \label{fig:fitke2g}
\end{figure}

For the measurement of the differential width, Eq. \ref{eq:rgamma}, we boost $E_\gamma(\mathrm{lab})$ 
to the kaon rest frame ($E_\gamma$) and perform independent fits for five $E_\gamma$ bins between 10 MeV 
and the kinematic limit, as defined in Table \ref{tab:countke2g}.
For each $E_\gamma$ bin, we are able to extract the number of \kedg\ events with $p_e>200$ MeV. 
Because of limited statistics, the counting is done combining the kaon charges.
Results are listed in Table \ref{tab:countke2g}. The total \kedg\ count, with $E_\gamma>10$ MeV and 
$p_e>200$ MeV,  is 1484$\pm$63 events.
\begin{table*}[ht]
  \begin{center}
    \begin{tabular}[c]{|c|c|c|c|c|c|}\hline
   $E_\gamma$ (MeV)&10 to 50&50 to 100&100 to 150&150 to 200&200 to 250\\\hline
   Signal counts &  55$\pm$16 &  219$\pm$24 & 463$\pm$32 & 494$\pm$38 & 253$\pm$26 \\
   $\chi^2/$ndf  &   80/66    &  141/105    &  87/106    & 100/106    & 116/102    \\      \hline
    \end{tabular}
  \end{center}
  \caption{Fit results for the number of \kedg\ events with $p_e>200$ MeV, in five $E_\gamma$ energy bins.}
  \label{tab:countke2g}
\end{table*}
Fig. \ref{fig:fitke2g} shows the sum of the fit results on all of the
$E_\gamma$ bins, projected onto the $\Delta E/\sigma$ axis for the signal region (top), defined as 
$m_\ell^2<8000$ MeV$^2$ or $14000<m_\ell^2<20000$ MeV$^2$, and for the background region (bottom). 
In the latter, \kmd\ dominate the region $8000<m_\ell^2<14000$ MeV$^2$,
while \ket\ dominate the region above $20000$ MeV$^2$ (see Fig.~\ref{fig:mlep_g}).

\section{Efficiency}
\label{sec:cs}
The ratios of \ked\ to \kmd\ and \kedg\ to \kmd\ efficiencies
are evaluated with MC and corrected for possible differences between data and MC, using control samples.
We evaluate data-MC corrections separately for each of the following analysis steps: 
decay point reconstruction (kink), quality cuts, cluster-charged particle association; for \kedg\ events, 
the efficiency for selection of a photon cluster is added, too. 
For each step, the correction is defined as the ratio of data and MC efficiencies measured on the 
control sample, each folded with the proper kinematic spectrum of \ked\ (or \kmd) events.

Decay point reconstruction efficiencies are evaluated using pure samples of \kmd\ and
\ket; these are tagged by the identification of the two-body decay, \kmd\ or $K \to \pi \pi^0$ (\kpd), 
of the other kaon~\cite{kchl3} and selected with tagging and EMC information only, 
without using tracking.  

A 99.5\% pure $\kmd^{\pm}$ sample is obtained with $K^\mp$ tagging plus one and only one EMC cluster with 
energy $E>90$ MeV, {\em not due} to the
tagging kaon decay products. The properties of the selected \kmd\ event are evaluated using time
and position of the cluster and the kaon momentum obtained from the
tagging (with 1\% resolution).
The muon momentum and the decay point position are determined a priori,
without using the kaon and electron tracks,
with a resolution of about 5 MeV and about 2 cm, respectively.
The tracking efficiency is determined as a function of the decay point position and the decay angle, by
counting the number of events in which a kink is reconstructed out of the number
of \kmd\  candidate events.

$\ket^{\pm}$ decays are selected in $K^{\mp}$ tagged events first detecting the photons from $\pi^0$ decay 
with time of flights consistent with a single point in the tagged kaon 
track obtained from the tagging kaon. Second, a third cluster with energy, time, and position consistent
with the expectation from a \ket\ decay is selected.
The electron momentum and the kaon decay point are determined a priori
with a resolution of \ab20 MeV (dominated by the measurement of $\pi^0$ momentum) and \ab2 cm, 
respectively.

The corrections to MC efficiencies range between 0.90 and 0.99 depending on the decay
point position and on the decay angle. The simulation is less accurate in case of
overlap between lepton and kaon tracks, and with decays occurring close
to the inner border of the fiducial volume.

Samples of $K_L(e3)$, $K_L(\mu 3)$, and \kmd\ decays with a purity of
99.5\%, 95.4\%, and 100.0\% respectively, are used to evaluate lepton cluster
efficiencies. These samples are selected using tagging and DC information only,
without using calorimeter, see Refs. \citen{klsl,ffe,ff}.
The efficiency has been evaluated as a function of
the particle momentum separately for barrel and endcap. The
correction to MC efficiencies ranges between 0.98 and 1.01 depending on the momentum and
on the point of impact on the calorimeter.

The single-photon detection efficiency for data and MC is evaluated
as a function of photon energy using \kpd\ events, in which one of the two photons
from $\pi^0$ decay is identified, allowing an a priori determination of the position and of
the energy of the second one. The average correction factor to MC efficiency is
$\sim$0.98.

The trigger efficiency has been evaluated solely from data. The probabilities
$\epsilon_\mathrm{EMC}^\mathrm{TRG}$ ($\epsilon_\mathrm{DC}^\mathrm{TRG}$)
for the EMC (DC) trigger condition to be satisfied in a DC-triggered
(EMC-triggered) event are evaluated in \ked-enriched
and \kmd-pure samples. The efficiency for the logical OR of the EMC and
DC trigger conditions is given by $\epsilon_\mathrm{EMC}^\mathrm{TRG}+
\epsilon_\mathrm{DC}^\mathrm{TRG}-
\epsilon_\mathrm{EMC}^\mathrm{TRG}\times\epsilon_\mathrm{DC}^\mathrm{TRG}$,
and it is $\sim0.99$ for both \ked\ and \kmd, with a ratio
$\epsilon^\mathrm{TRG}(\ked)/\epsilon^\mathrm{TRG}(\kmd)=0.9988(5)$.
A possible bias on the previous result due to correlation between EMC and DC
triggers is also taken into account, which is evaluated to be $0.997(1)$
using MC simulation.

The event losses induced by the cosmic veto applied at the trigger level
and by the background rejection filter applied offline 
(FILFO) are evaluated from samples of downscaled events,
in which the veto conditions are registered but not enforced.
The ratio of \ked\ to \kmd\ efficiencies are $1.0013(2)$ and
$0.999(4)$ for cosmic veto and FILFO, respectively.
The statistical error due to the FILFO correction is $0.4\%$, and
dominates the total uncertainty in trigger, cosmic veto, and FILFO corrections.

\def\x{\times}
\def\dsten{$\delta(R_{10})$}  \def\dsgam{$\delta(R_{\gam})$}

\section{Systematic errors}
\label{sec:syst}

The absolute values of all of the systematic uncertainties on $R_{10}$ and $R_{\gamma}$, the integral
of Eq. \ref{eq:rgamma} for $E_\gamma > 10$ MeV, are listed in Table \ref{tab:syske2}; as a comparison, 
the statistical uncertainty is reported as well. All of the sources of systematic error 
are discussed below.
\begin{table}[ht!]
  \begin{center}
    \begin{tabular}[c]{|cccc|}\hline
                          & &\vp \dsten$\x10^5$ & \dsgam$\x10^5$ \\ \hline
    \multicolumn{2}{|c}{Statistical error}                 & 0.024 & 0.066 \\ \hline
    \multicolumn{2}{|c}{Systematic error}                  &       &       \\
            Counting: &    fit                             & 0.007 & 0.004 \\
                      &     DE                             & 0.005 &  -    \\
            Efficiency:& kink                              & 0.014 & 0.009 \\
                & trigger                                  & 0.009 & 0.006 \\
                & $e, \mu$ cluster                         & 0.005 & 0.003 \\
                & $\gamma$ cluster                         & -     & 0.003 \\
               \multicolumn{2}{|c}{Total systematic error} & 0.019 & 0.013 \\ \hline
    \end{tabular}
  \end{center}
  \caption{Summary of statistical and systematic uncertainties on the measurements of 
           $R_{10}$ and $R_{\gamma}$.}
  \label{tab:syske2}
\end{table}

To minimize possible biases on \ked\ event counting due to the
limited knowledge of the momentum resolution, we used \kmd\ data to carefully tune the MC response on the
tails of the $m_\ell^2$ distribution. This has been performed in sidebands of the \nn\ variable, to
avoid bias due to the presence of \ked\ signal.
Similarly, for the \nn\ distribution the EMC response in the MC has been tuned at the level of single cell,
using $K_{\ell 3}$ data control samples. 
Residual differences between data and MC \ked\ and \kmd\ \nn\ shapes have been corrected by using 
the same control samples.
Finally, to evaluate the systematic error associated with these procedures, 
we studied the variation of the results
with different choices of fit range, corresponding to a change of overall purity from
$\sim75\%$ to $\sim10\%$, for $\ke(\gamma)$ with $E_\gamma<10$ MeV, and from
$\sim31\%$ to $\sim10\%$, for $\ke(\gamma)$ with
$E_\gamma>10$ MeV and $p_e>200$ MeV.  The results are stable within
statistical fluctuations.  A systematic uncertainty of $\sim0.3\%$ for both
$R_{10}$ and $\mathrm{d}R_\gamma/\mathrm{d}E_\gamma$, independently on
$E_\gamma$, is derived by scaling the uncorrelated errors so that the reduced $\chi^2$ value equals unity
(see also Table \ref{tab:syske2}).

\ked\ event counting is also affected by the uncertainty on $f_\mathrm{DE}$, the fraction
of \ked\ events in the fit region which are due to DE process. 
This error has been evaluated by repeating the measurement of
$R_{10}$ with values of $f_\mathrm{DE}$ varied within its uncertainty, 
which is $\sim 4\%$ according to our measurement
of the \kedg\ differential spectrum (Sects. \ref{sec:kedg} and \ref{sec:results}).
Since the $m_\ell^2$ distributions for \kedg\ with $E_\gamma<10$ MeV and with $E_\gamma>10$ MeV
overlap only partially, the associated fractional variation on $R_{10}$ is reduced:
the final error due to DE uncertainty is 0.2\% (Table \ref{tab:syske2}).

Different contributions to the systematic uncertainty on
$\epsilon_{e2}/\epsilon_{\mu2}$ are listed in Table \ref{tab:syske2}.
These errors are do\-mi\-na\-ted by the statistics of the control samples used to
correct the MC evaluations.  In addition, we studied the variation
of each correction with modified control-sample selection criteria. We
found neglible contributions in all cases but for the kink
and quality cuts corrections, for which the bias due to the control-sample selection
and the statistics contribute at the same level.

The total systematic error is $\sim0.8\%$ for both $R_{10}$ and $R_\gamma$ measurements, to be compared
with statistical accuracies at the level of $\sim 1\%$ and $\sim 4\%$, respectively.
As a further cross-check on the results, and particularly on the criteria adopted to obtain the 
data/MC corrections,
we measured with the same analysis method the ratio $R_{\ell 3}=\mathrm{\Gamma}(\ket)/\mathrm{\Gamma}(\kmt)$.
We found $R_{\ell 3}=1.507\pm0.005_\mathrm{stat}$ and $R_{\ell 3}=1.510\pm0.006_\mathrm{stat}$ for $K^+$ and $K^-$.
These results agree within the quoted accuracy 
with the value expected from the world-average form-factor slope measurements \cite{flavia},
$R_{\ell 3}=1.506\pm0.003$.

\section{Results and interpretation}
\label{sec:results}

\subsection{\mbo $R_K$ and lepton-flavor violation}
The number of $\ke(\gamma)$ events with $E_\gamma<10$ MeV,
the number of $\km(\gamma)$ events, the ratio of \ked\ to \kmd\ efficiencies
and the measurement of $R_{10}$ are given in Table \ref{tab:fitresultske2} for
$K^+$, $K^-$ and both charges combined.
$K^+$ and $K^-$ results are consistent within the statistical error. The systematic uncertainty is
common to both charges.
\begin{table*}[ht!]
  \begin{center}
    \begin{tabular}[c]{|c|c|c|c|c|}\hline
        & N($K_{e2}$) &   N($K_{\mu2}$)   &  $\epsilon_{e2}/\epsilon_{\mu2}$ & $R_{10}$  \\ \hline
\vp$K^+$ & 6348 $\pm$ 92 $\pm$ 23&$2.878\times10^8$&$0.944\pm0.003\pm0.007$&$(2.336\pm0.033\pm0.019)\times10^{-5}$\\
   $K^-$ & 6064 $\pm$ 91 $\pm$ 22&$2.742\times10^8$&$0.949\pm0.002\pm0.007$&$(2.330\pm0.035\pm0.019)\times10^{-5}$\\
   $K^\pm$&12412 $\pm$ 129 $\pm$ 45&$5.620\times10^8$&$0.947\pm0.002\pm0.007$&$(2.333\pm0.024\pm0.019)\times10^{-5}$\\ 
\hline
    \end{tabular}
  \end{center}
  \caption{Number of \ked\ and \kmd\ events, efficiency ratios and results for $R_{10}$
   for $K^+$, $K^-$, and both charges combined; first error is statistical, second one is systematic.}
  \label{tab:fitresultske2}
\end{table*}

\begin{figure}[ht!]\centering
  \includegraphics[width=0.8\linewidth]{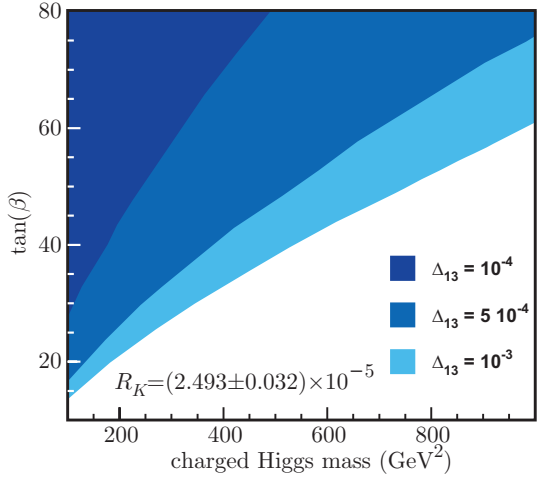}
  \caption{Excluded regions at 95\% C.L. in the plane
 $M_H$--$\tan \beta$ for $\Delta_R^{31}=10^{-4}, 5\times 10^{-3},10^{-3}$.}
  \label{fig:rkmssm}
\end{figure}

To compare the $R_{10}$ measurement with the inclusive $R_K$ prediction from SM, we take
into account the acceptance of the 10 MeV cut for IB, Eq. \ref{eq:R10new}. We obtain:
\be
R_K = (2.493\pm0.025_\mathrm{stat}\pm0.019_\mathrm{syst})\times 10^{-5},
\ee
in agreement with SM prediction of Eq. \ref{eq:rksm}. 
In the framework of MSSM with lepton-flavor violating (LFV)
couplings, $R_K$ can be used to set constraints in the space of relevant parameters, 
using the following expression \cite{masiero}:
\be
\label{eq:rkmssm}
  R_K=R_K^{\mathrm{SM}}\times\left[1+ \left(\frac{m_K^4}{m_H^4}\right)
  \left(\frac{m^2_\tau}{m^2_e}\right)
  \left|\Delta_R^{31}\right|^2\tan^6\beta\right], \ee
where $M_H$ is the charged-Higgs mass, $\Delta_R^{31}$ is the effective $e$-$\tau$
coupling constant depending on MSSM parameters, and $\tan\beta$ is the
ratio of the two Higgs superfields vacuum expectation values.
The regions excluded at 95\% C.L. in the plane $M_H$--$\tan\beta$ are shown
in Fig. \ref{fig:rkmssm} for different values of the effective LFV
coupling $\Delta_R^{31}$.

\subsection{\mbo Measurement of $\mathrm{d}R_\gamma/\mathrm{d}E_\gamma$}
\begin{figure}[ht!]\centering
\includegraphics[width=0.85\linewidth]{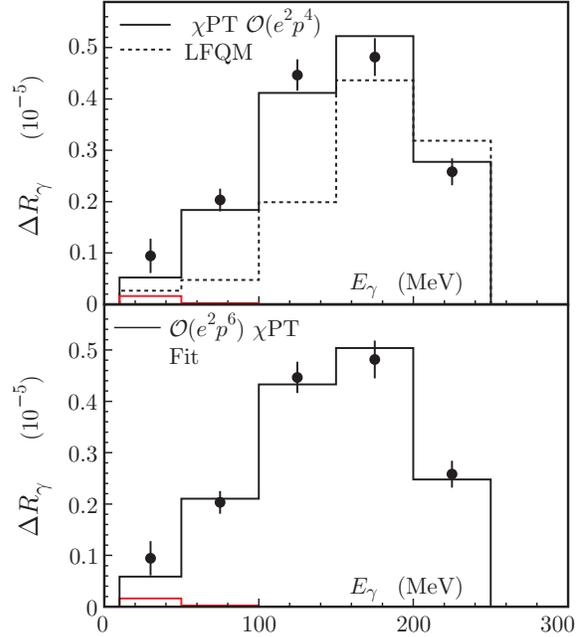}
\caption{$\Delta R_\gamma=\left[1/\Gamma(K_{\mu2})\right]\x[\dif\Gamma(K_{e2\gamma}/\dif E_\gamma]$ vs $E_\gamma$.
On top data (black dots) are compared to \ChPT\  predictions at ${\mathcal O}(e^2p^4)$ and to the LFQ model, see text.
At the bottom data are fitted to \ChPT\ at ${\mathcal O}(e^2p^6)$. The IB contribution is shown (red line).}
\label{fig1-2}
\end{figure}

Results on the differential spectrum are given in Table \ref{tab:reske2g}.
For each $E_\gamma$ bin we measure $\Delta$\erreg, 
the integral of $\mathrm{d}R_\gamma/\mathrm{d}E_\gamma$ over the bin width.
\begin{table}[ht]
\begin{center}
\begin{tabular}[c]{|c|c|c|}\hline
\vp$E_\gamma$ (MeV)&$\epsilon$(e2)/$\epsilon$($\mu$2)&$\Delta$\erreg\ ($10^{-6}$)\\ \hline
 10 to 50 &0.104$\pm$0.003&0.94$\pm$0.30$\pm$0.03\\
 50 to 100&0.192$\pm$0.001&2.03$\pm$0.22$\pm$0.02\\
100 to 150&0.184$\pm$0.001&4.47$\pm$0.30$\pm$0.03\\
150 to 200&0.183$\pm$0.001&4.81$\pm$0.37$\pm$0.04\\
200 to 250&0.174$\pm$0.002&2.58$\pm$0.26$\pm$0.03\\ \hline
\end{tabular}
  \end{center}
  \caption{$\mathrm{d}R_\gamma/\mathrm{d}E_\gamma$ results.
           Most of the efficiency ratio error is common to all energy bins.}
  \label{tab:reske2g}
\end{table}
In Fig. \ref{fig1-2} top,
our measurements are compared to the prediction from \ChPT\ at
${\mathcal O}(p^4)$ \cite{bijnens} and from the Light Front Quark model (LFQ) of  Ref. \citen{Korea}. 
Integrating over $E_\gamma$ from 10 MeV to 250 MeV, we obtain:
\be
R_\gamma = (1.483 \pm 0.066_\mathrm{stat}\pm 0.013_\mathrm{syst})\times 10^{-5},
\ee
in agreement with the prediction $R_\gamma=1.447\times10^{-5}$, which is obtained using the
values for the effective couplings ($V$ and $A$) from ${\mathcal O}(e^2p^4)$ \ChPT\ \cite{bijnens}
and using world-average values for all of the other relevant parameters. The $R_\gamma$ prediction includes a 1.32(1)\%
contribution from IB. This result confirms within a 4\% error the amount of DE component in our MC.

The comparison of the measured spectrum with the \ChPT\ prediction
shown in Fig. \ref{fig1-2} top suggests the presence of a form
factor, giving a dependence of the effective couplings on the transeferred
momentum, $W^2=M_K^2(1-x)$, as predicted by \ChPT\ at ${\mathcal O}(e^2p^6)$ \cite{Korea}.
The form-factor parameters are obtained by
fitting the measured $E_\gamma$ distribution with the theoretical differential
decay width given in Eq. \ref{eq:dGdeDE}, with the vector effective
coupling expanded at first order in $x$:
$V=V_0(1+\lambda(1-x))$. The axial effective coupling $A$ is assumed
to be independent on $W$ as predicted by \ChPT\ at ${\mathcal O}(e^2p^6)$ \cite{Korea}.
The small contribution from DE$^-$ transition to
our selected events does not allow a fit to the related $V-A$
component. Therefore, in the fit $V_0-A$ is kept fixed at the expectation from \ChPT\ at $O(e^2p^4)$,
while $V_0+A$ and $\lambda$ are the free parameters.
The result of this fit is shown in Fig. \ref{fig1-2} bottom. We obtain:
\begin{eqnarray*}
  V_0+A   & = & 0.125\pm0.007_\mathrm{stat}\pm0.001_\mathrm{syst},\\
  \lambda & = & 0.38\pm0.20_\mathrm{stat}\pm0.02_\mathrm{syst},
\end{eqnarray*}
with a correlation of -0.93 and a $\chi^2/\mbox{ndof}=1.97/3$. 
Our fit confirms at $\sim 2\sigma$ the presence of a slope in the vector
form factor, in agreement with the value from \ChPT\ at $O(e^2p^6)$, $\lambda\sim0.4$.

\section{Conclusions}
We have performed a comprehensive study of the process \kedg. We have measured
the ratio of \kedg\ and \kmd\ widths for photon energies smaller than 10 MeV, without photon detection
requirement.  We find:
\be R_{10}=(2.333\pm0.024_\mathrm{stat}\pm0.019_\mathrm{stat})\times 10^{-5}.
\ee
From this result we derive the inclusive ratio $R_K$ to be compared with the SM prediction:
\be R_K=(2.493\pm0.025_\mathrm{stat}\pm0.019_\mathrm{syst})\times 10^{-5},
\ee
in excellent agreement with the SM prediction
\be R_K= (2.477\pm0.001)\times 10^{-5}.\ee
Our result improves the accuracy with which $R_K$ is known by a factor of 5 with respect to the
present world average and allows severe constraints to be set
on new physics contributions in the MSSM with lepton flavor violating couplings as shown
in Fig. \ref{fig:rkmssm}.

To obtain the value of $R_K$ from the measurement of $R_{10}$ knowledge of radiative effects
is required for both inner bremsstrahlung and direct emission. The latter is important for the
helicity suppressed \ke\ decay but is not precisely known nor the differential width has ever been measured.
We have therefore measured the differential decay width for \kedg\ as a function of
$E_\gamma$, normalized to \kmd, in the
momentum region $p_e>200$ MeV, in the kaon rest frame. Our result for the direct emission width
is in agreement with the expectation from \ChPT\ and gives an indication of the presence
of ${\mathcal O}(e^2p^6)$ contributions.

\vspace{1.cm}
{\it Acknowledgements}
We thank the \DAF\ team for their efforts in maintaining low background running 
conditions and their collaboration during all data-taking. 
We want to thank our technical staff: 
G.~F.~Fortugno and F.~Sborzacchi for their dedicated work to ensure an
efficient operation of the KLOE Computing Center; 
M.~Anelli for his continuous support to the gas system and the safety of the detector; 
A.~Balla, M.~Gatta, G.~Corradi and G.~Papalino for the maintenance of the electronics;
M.~Santoni, G.~Paoluzzi and R.~Rosellini for the general support to the detector; 
C.~Piscitelli for his help during major maintenance periods.
This work was supported in part by EURODAPHNE, contract FMRX-CT98-0169; 
by the German Federal Ministry of Education and Research (BMBF) contract 06-KA-957; 
by the German Research Foundation (DFG),'Emmy Noether Programme', contracts DE839/1-4.

\end{document}